\documentclass[a4paper,11pt]{article}
\pdfoutput=1 % if your are submitting a pdflatex (i.e. if you have
\usepackage{jinstpub} % for details on the use of the package, please
\title{\boldmath Precision timing calorimetry with the CMS HGCAL}

\author{Artur Lobanov}
\affiliation{Laboratoire Leprince-Ringuet, \'Ecole polytechnique, IN2P3/CNRS\\ 91128 Palaiseau, France}

\emailAdd{artur.lobanov@cern.ch}

\abstract{
The existing CMS endcap calorimeters will be replaced with a High Granularity Calorimeter (HGCAL) for operation at the High Luminosity (HL) LHC. Radiation hardness and excellent physics performance will be achieved by utilising silicon pad sensors and SiPM-on-scintillator tiles with high longitudinal and transverse segmentation. 
One of the major challenges of the HL-LHC will be the high pileup environment, with interaction vertices spread not only in position, but also in time. In order to efficiently reject particles originating from pileup, precision timing information of the order of 30\,ps will be of great benefit.
In order to meet such performance goals, the HGCAL will provide timing measurements for individual hits with signals above 12\,fC (equivalent to 3-10\,MIPs), such that clusters resulting from particles with $p_T$ > 5\,GeV should have a timing resolution better than 30\,ps. Given the complexity and size of the system, this poses a particular challenge to the readout electronics as well as to the calibration and reconstruction procedures.
We present the challenges for the front-end electronics design, results from prototype tests in laboratory and beam environments, as well as anticipated timing performance from simulation.
}

\keywords{Calorimeters, Front-end electronics for detector readout, Precision timing}

\collaboration[c]{on behalf of the CMS collaboration}

\proceeding{CHEF2019: Calorimetry for the High Energy Frontier\\
  25-29 November 2019\\
  Fukuoka, Japan}

\usepackage{lineno}
\begin{document}
\maketitle
\flushbottom

%CMS~\cite{Chatrchyan:2008aa} is great.

\section{Precision timing for the HL-LHC}

The next phase of the already successful Large Hadron Collider (LHC), called the High-Luminosity LHC (HL-LHC)  is foreseen to start operation in 2027. Within a 10 year time period the HL-LHC will provide about ten times the LHC dataset. This rate will require an instantaneous luminosity three to four times higher than presently available. Such an exceptional environment poses great challenges to the experimental detectors, which were designed for the current LHC phase.

In order to cope with the more challenging environment and mitigate effects of radiation damage, the CMS collaboration~\cite{Chatrchyan:2008aa} is planning an upgrade of its detector systems and infrastructure. 
Due to the extremely harsh conditions in the forward region, the present endcap calorimeters will be replaced by the new High Granularity Calorimeter (HGCAL)~\cite{TDR, FS}. 
Finely segmented silicon sensors and scintillator tiles with silicon photomultiplier  readout will cover 50 longitudinal layers in order to provide high granularity information for physics objects reconstruction.

A grand challenge of the HL-LHC will be the large pileup with up to an average of 200 simultaneous proton collisions. The interaction vertices will be spread out by $\pm 50$\,mm along the beam in distance, as well as in time, about $\pm 150$\,ps. These characteristics imply that timing measurements with resolutions in the 30\,ps range will greatly enhance pileup mitigation.
In order to meet these performance goals, HGCAL will provide timing measurement for individual hits with signals above 12\,fC (equivalent to 3-10\,MIPs), such that clusters with $p_T > 5$ \,GeV should have a timing resolution better than 30\,ps.

\section{Timing use in reconstruction}
\label{sec:reco}
Extensive studies using full simulations  were performed  to evaluate the use of timing information in the reconstruction of  physics objects within HGCAL (cf. chapter 5.5 in \cite{TDR}).
Based on electronics simulations, the single channel resolution was assumed to be:
% have a noise term of 1.5\,ns/Q and a constant term of 20\,ps.
$$
\sigma_t = \sigma _{noise} \oplus  \sigma _{floor},
\textrm{where } \sigma _{noise} = \frac{A}{S/N},
$$
with a noise term $A$ of 1.5\,ns/fC, a constant term $\sigma _{floor}$  of 20\,ps.  $S/N$ denotes the signal-over-noise ratio, whereas the symbol $\oplus$ indicates quadratic summation.

Given such a noise term, single-hit timing information may not allow to discriminate hits from pileup objects.
However, the large granularity results in a high hit multiplicity, and therefore the best timing performance is expected from the combination of all cells from within the same particle shower.  

With the studies presented in the HGCAL TDR, it was shown that rejecting hits in the tails provides robust  pileup rejection. Further, taking only cells within 2\,cm of the shower axis, an efficiency of 100\,\% for photons is achievable with a resolution of 20\,ps for $p_T > 2$\,GeV. For neutral kaons, $K^0_L$, with $p_T > 5$\,GeV the resolution is  below 30\,ps with a 90\,\% efficiency.

\section{Electronics requirements}

Precision timing in such a complex, high granularity detector is a great challenge for the electronics system of the whole experiment. Therefore, very strict requirements on the timing performance is imposed on all components, starting from the sensitive elements up until the clock distribution system.
%Single HGCAL sensor timing performance evaluated in 2016 beam tests [JINST 13 (2018) P10023]
The timing performance of single HGCAL prototype silicon sensors has been evaluated in a dedicated beam test campaign, see Ref.~\cite{TB2016}. After irradiation up to a fluence of  $10^{16} \textrm{cm}^{-2}  \textrm{ MeV neq}$ the timing resolution constant term for single channels was found to be about 20\,ps. The clock distribution system is being designed to have a reduced jitter ($<$ 15\,ps).
Thus, the already very challenging front-end ASIC HGCROC~\cite{HGCROC} is key to the overall timing performance, since it is responsible for the precise measurement of the incoming physics signals. 

\begin{figure}[htbp]
	\centering
	\includegraphics[width=0.49\linewidth]{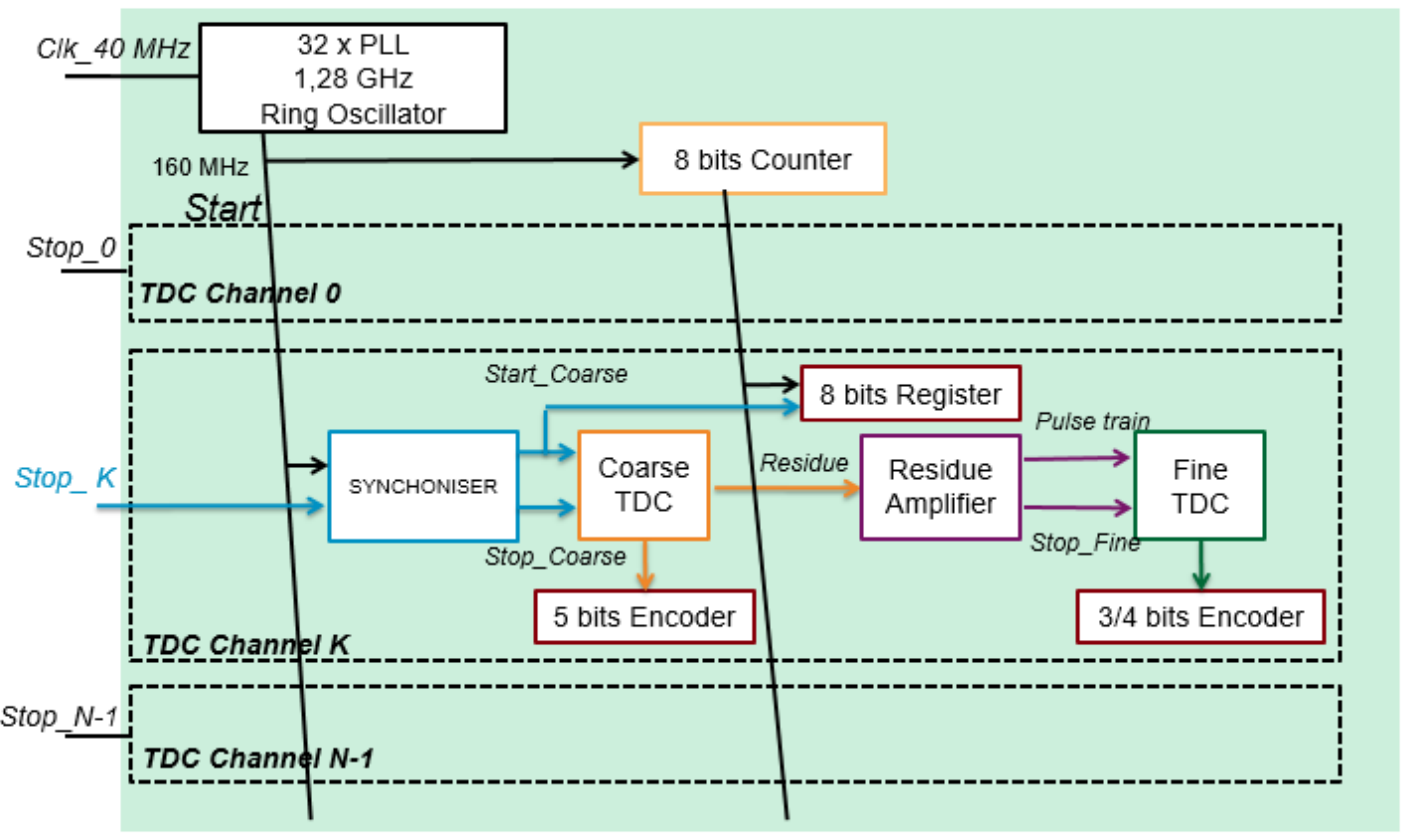}
	\includegraphics[width=0.5\linewidth]{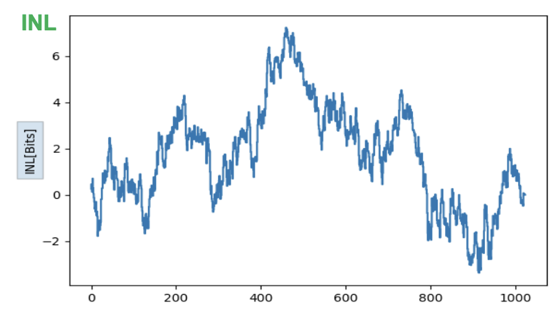}
	\caption{Left: architecture of the TDC block in the HGCROC. Right: integral non-linearity (INL) of the uncalibrated TDC measurement with respect to the 10-bit TDC code.}
	\label{fig:hgcroc_tdc}
\end{figure}

The HGCROC features a Time-of-Arrival (TOA) block based on a constant-threshold discriminator and a 3-stage TDC (Fig.~\ref{fig:hgcroc_tdc}). The first stage uses a 2 bit gray counter on the 160 MHz clock of the chip internal phase-locked loop (PLL), hence with a least significant bit (LSB) of 6.25\,ns.
Further, a coarse time-to-digital converter (TDC) with a classical delay-locked loop (DLL) measures the time between the last 40 MHz clock edge and the trigger signal with an LSB of 195\,ps. 
Finally, a fine TDC with an LSB of 24.4\,ps uses a residue integrator based on a DLL line.
The architectural advantages of such a design are high-speed conversion, low power consumption, and large time range due to the global counter. This design also ensures the same performance under temperature and process variations.

Performance measurements are currently being performed with HGCROC prototypes.
For the TDC block, the differential non-linearity is within 1 LSB, whereas the raw integral non-linearity (INL) is about 10 LSB peak-to-peak. After INL calibration, the TDC time resolution is found to be about 10\,ps. The full analog and digital TOA chain is reaching a resolution of about 50\,ps satisfying the specification.

\section{Prototype timing performance in beam test }
The performance and feasibility of the HGCAL design have been validated in several beam test campaigns in the years from 2016 until 2018. A major beam test has taken place in October 2018 at the CERN SPS H2 beamline, featuring almost 100 silicon prototype modules~\cite{TB}. 

\begin{figure}[tb]
	\centering
	\includegraphics[width=0.54\linewidth]{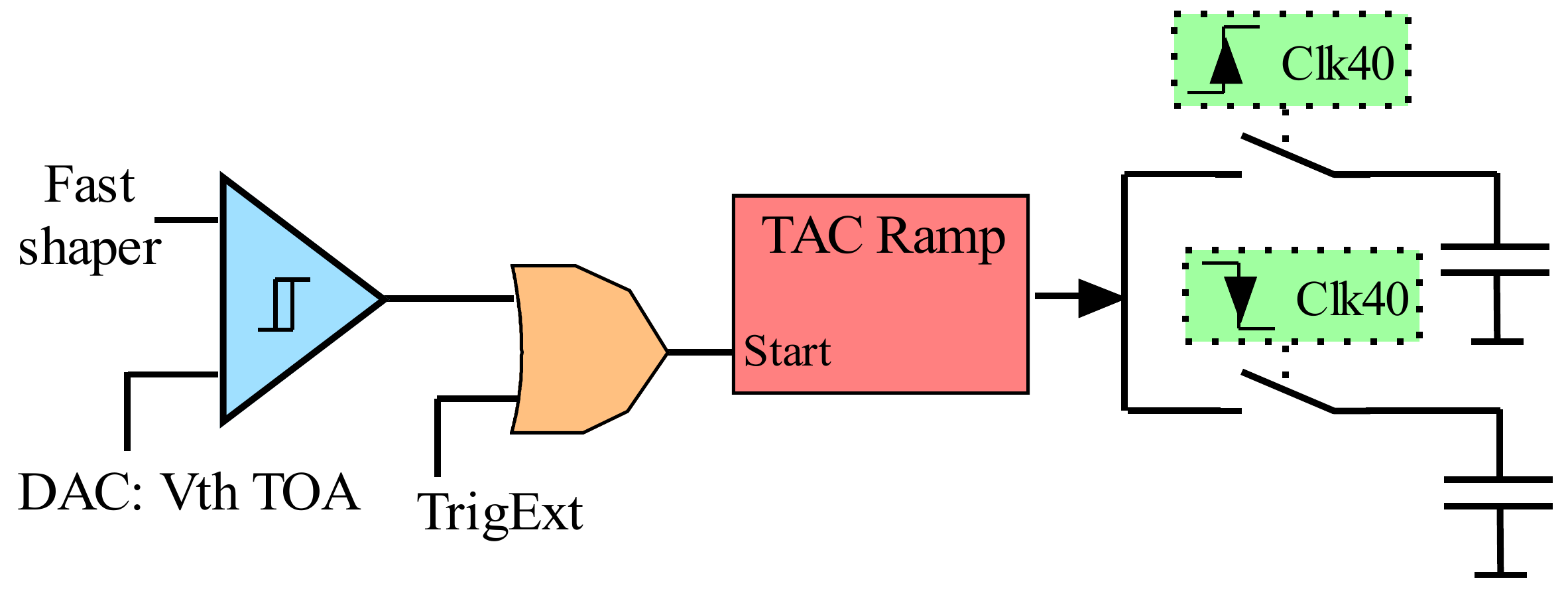}
	\includegraphics[width=0.45\linewidth]{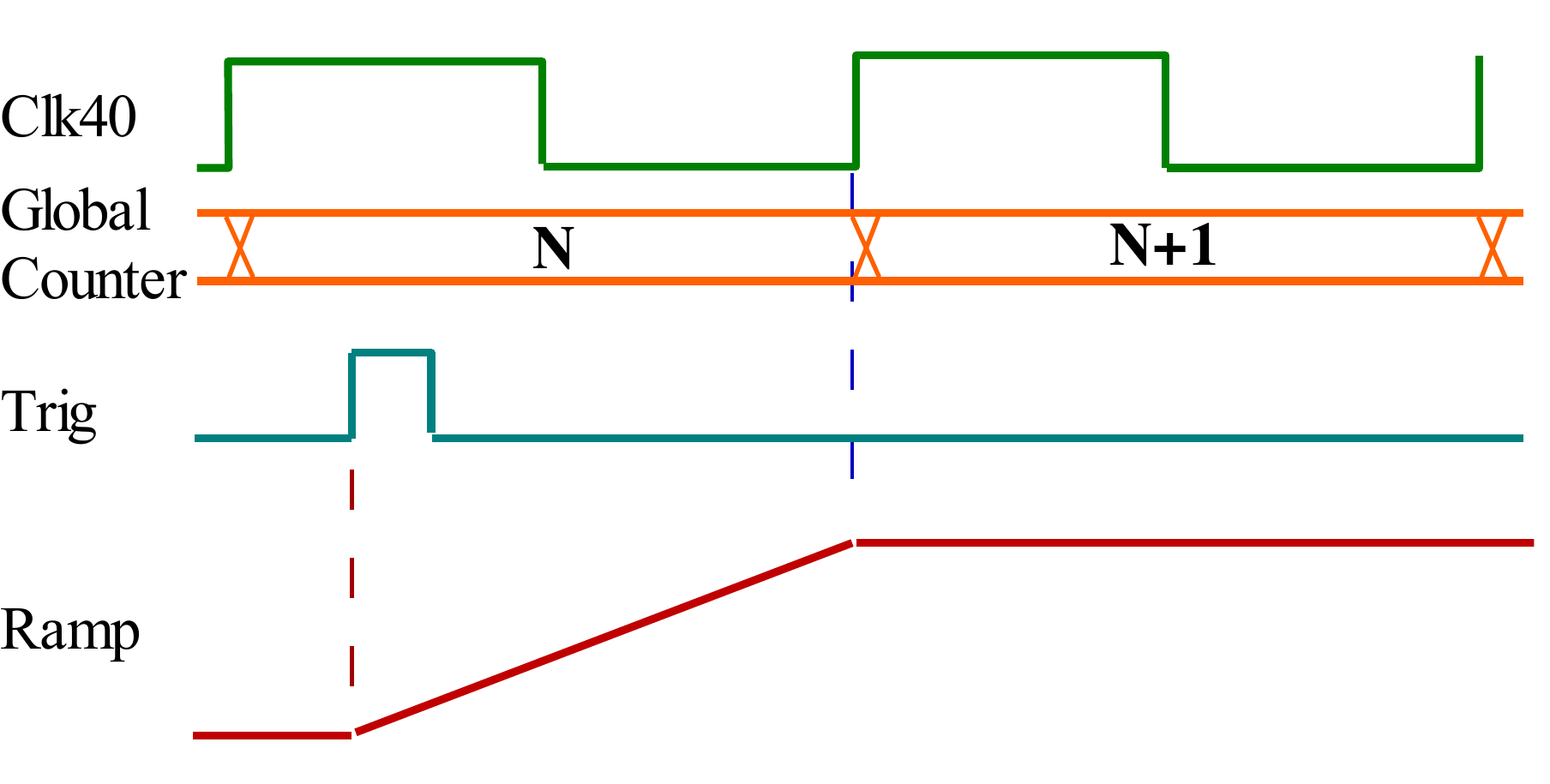}
	\caption{SKIROC2-CMS TOA analog blocks (left) and time-to-amplitude (TAC) ramp (right).}
	\label{fig:toa_block}
\end{figure}

One of the goals of these tests was to validate the the timing performance of the prototype.
For this purpose, the dedicated front-end readout ASIC SKIROC2-CMS~\cite{SK2cms} was equipped with a Time-of-Arrival measurement circuit (Fig.~\ref{fig:toa_block}).
The output of the preamplifier is fed into a CRRC fast shaper, followed by a threshold discriminator which starts a voltage ramp (Time-to-Amplitude converter, TAC) that is sampled at the edge of the 40 MHz system clock. 
The fast shaping time is set to 5\,ns,  and the discriminator threshold to an equivalent of about 40\,fC  for an optimal noise and timing performance. 
Due to the design specifics, the TAC ramp saturates after about 12\,ns causing a non-linearity in large TOA values. The measurements of the rising and falling clock edge, thus offset by 12.5\,ns, allow to reduce the impact of the non-linearity by taking a weighted average.

Measurements of the TOA resolution on a single-ASIC test board resulted in a constant term of 50\,ps, compatible with the specifications. In order to effectively asses the performance in the beam test, a MCP-PMT device with a 20\,ps resolution was used as a time reference.

%TOA from fast shaper (shaping time = 5ns) followed by discriminator  and Time-to-Amplitude converter (TAC = TDC)
%TAC based on voltage ramp stopped at clock edge (rising, falling)
%Ramp saturation causes non-linearity in the response
%TOA resolution measured to reach 50ps with single chip (chip specification!)
%Reference timing device: MCP-PMT (<20ps resolution)

\begin{figure}[htb]
	\centering
	\includegraphics[width=0.4\linewidth]{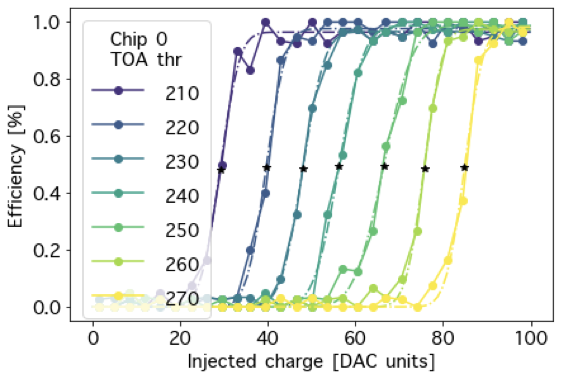}
	\qquad
	\includegraphics[width=0.35\linewidth]{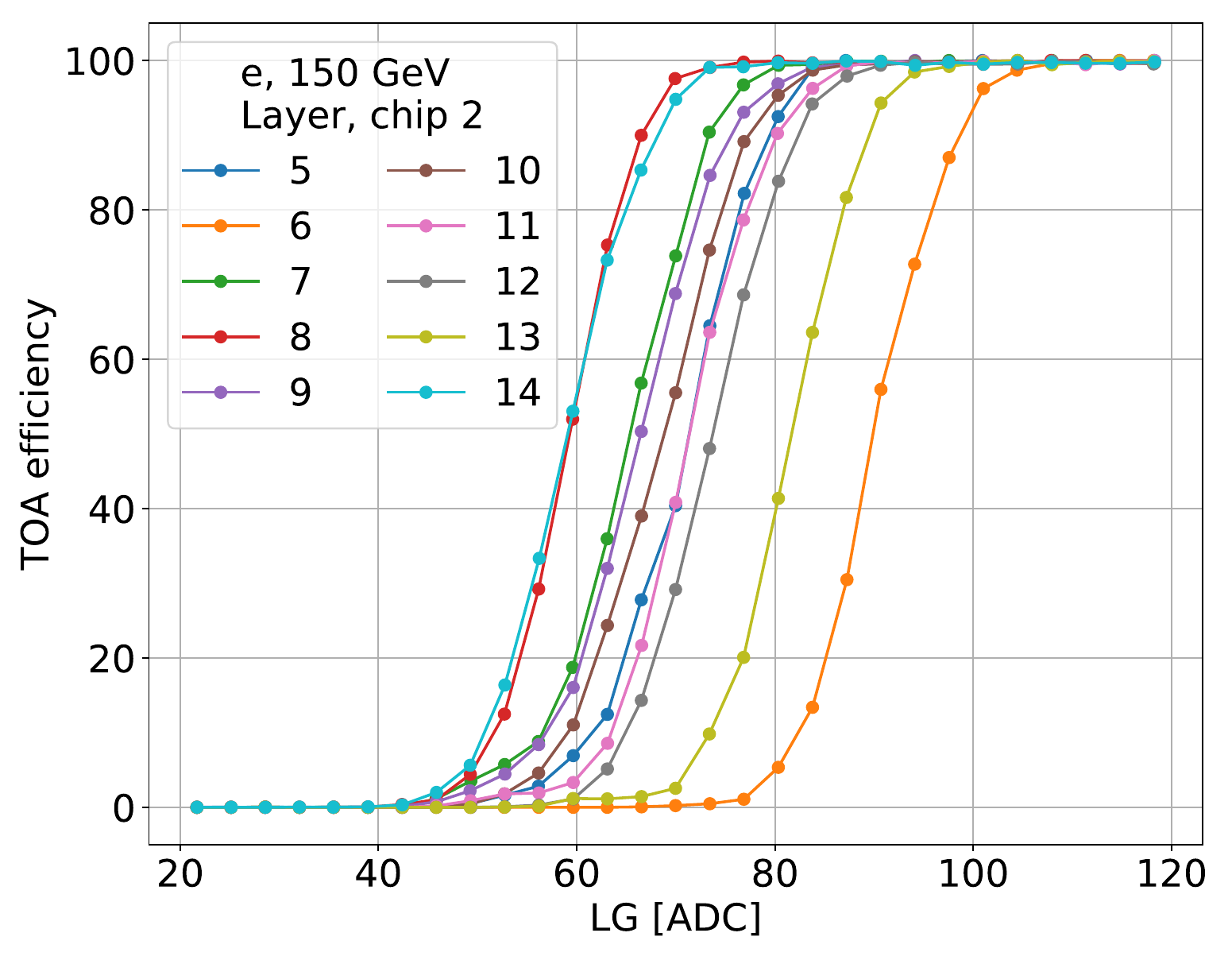}
	\caption{TOA turn-on curves for a threshold scan (left) and as measured in beam  data (right).}
	\label{fig:toa_scurves}
\end{figure}

After the initial large common-mode noise found in tests during 2017 was eliminated, the TOA threshold was optimised for  data taking in 2018.
Since noise can cause TOA inefficiencies, the thresholds were determined using the turn-on curve method with external pulse injection in laboratory environment (Fig.~\ref{fig:toa_scurves}). The thresholds were selected corresponding to 95\,\% TOA efficiency. Measurements using the beam test data showed that the effective TOA thresholds were in a range from 10 to 20 minimum-ionising particle (MIP) charge equivalent.

\begin{figure}[h]
	\centering
	\includegraphics[width=0.32\linewidth]{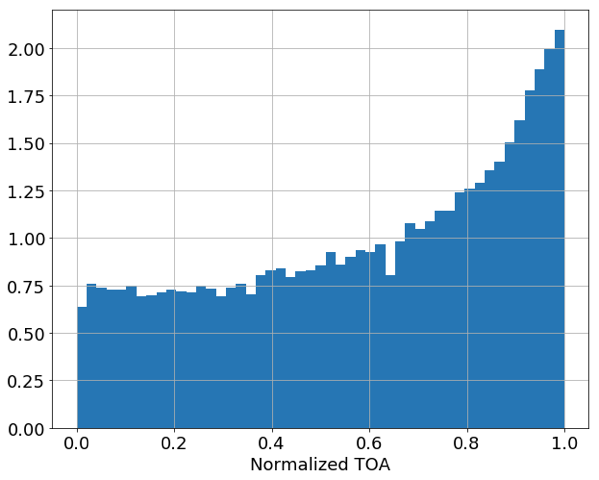}
	\includegraphics[width=0.31\linewidth]{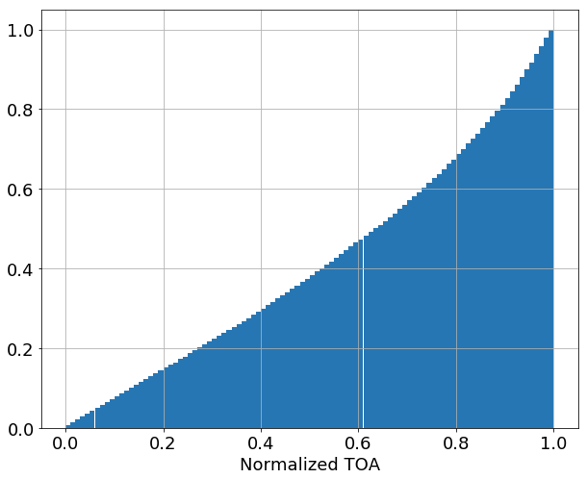}
	\includegraphics[width=0.3\linewidth]{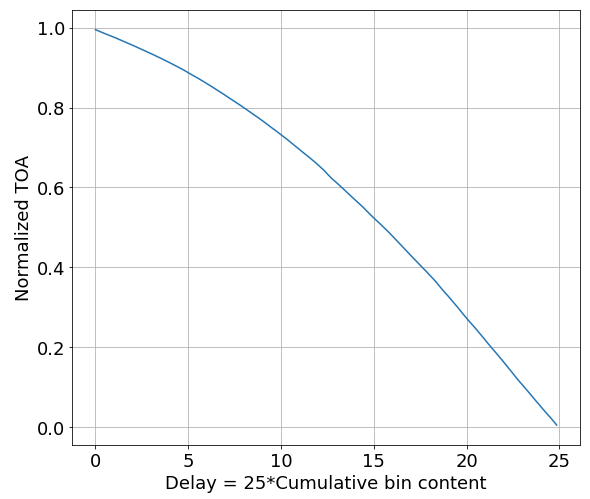}
	\caption{Distribution of the uncalibrated normalized TOA in beam test data (left), its cumulative distribution (middle), and the reconstructed TOA response curve for the true hit time (delay) from the CDF (right).}
	\label{fig:toa_calib1}
\end{figure}

A dedicated data-driven calibration method was developed in order to find the response curve of the TOA.
The method is based on the fact that the SPS beam is asynchronous with respect to the ASIC clock used for the TOA measurement. Therefore, the true hit time distribution is uniform in its 25\,ns range.
According to the probability integral transform\footnote{From probability theory: it states that if a random variable X has a continuous distribution for which the CDF is $F_X$, then the random variable defined as $F_X(X)$ has a standard uniform distribution.},
the variable constructed from the cumulative distribution function (CDF) of the TOA has a uniform distribution. Hence, this variable is proportional to the true hit time. Due to the normalization of the CDF, a multiplication by 25 is needed to obtain the unit in nanoseconds (Fig.~\ref{fig:toa_calib1}).
Note that for convenience, normalized TOA values in the range from 0 ro 1 are used.
%This notion allows the comparison of normalized TOA distributions between different channels and ASICs.
In addition, an inversion of the delay and TOA variables is performed to reflect the fact that
% larger TOA values mean earlier time (
the measurement is done against the following clock edge.
In such a representation the TAC non-linearity is visible for large TOA and low delay values (Fig.~\ref{fig:toa_calib1}, right).

%Exploiting: beam is asynchronous wrt the clock –> true TOA distribution is uniform in [0,25) ns interval
%Goal: restore TOA(delay) function given TOA distribution sampled from a uniform delay distribution
%Result: TOA histogram from TOA(delay) sampling with delay = U[0,1) in orange (similar to random number generation from p.d.f.!)
%Distribution coincides with derivative of delay(TOA):  
%Restore TOA(delay) from cumulative TOA distribution!

\begin{figure}[htb]
	\centering
	\includegraphics[width=0.4\linewidth]{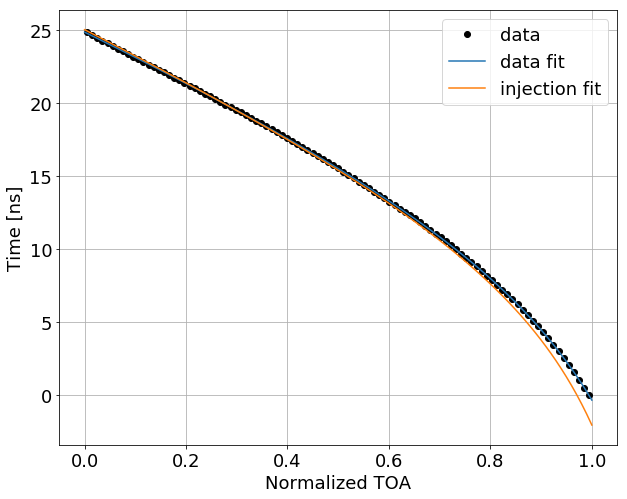}
	\qquad
	\includegraphics[width=0.3\linewidth]{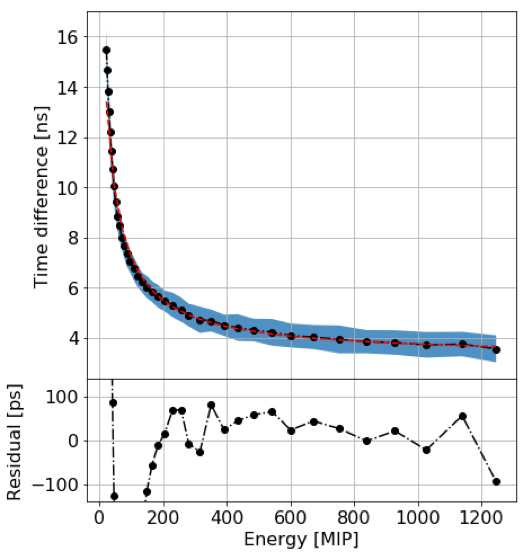}
	
	\caption{Left: comparison of the TOA response curve derived in beam data (black dots) and with external charge injection (dashed line). Right: TOA timewalk of a single channel with respect to the MCP-PMT reference.}
	\label{fig:toa_calib2}
\end{figure}

This data-driven calibration method was validated in several ways. 
The median of the raw TOA distribution corresponds to the 12.5\,ns point as can be exactly verified from the measurement of  both the rise and fall time.
%The beginning of the reconstructed TOA curve is compatible with a line with zero intercept, corresponding to the linear part of the TOA ramp.
The linear part of the TOA is compatible with an intercept of 25\,ns, which is the maximum range of the TOA.
Finally, a comparison of the calibration curves obtained from the data-driven method and independently from external charge injection shows a reasonable agreement for most of the range (Fig.~\ref{fig:toa_calib2}, left). 

%Applying the method in beam data:
%Normalise TOA ranges (~TOA pedestals)
%Construct cumulative histogram from TOA distribution
%Take histogram bin content as (normalised) true delay -> fit TOA(delay) dependency

%Verifying whether the obtained curve (and method) gives a correct result:
%Median of TOA distribution coincides with 12.5 ns point (= rise/fall transition)
%Beginning of the TOA vs delay is perfectly linear with p0 = 0
%Curve points and fitted function well compatible with injection fit \& MCP reference

%Performing TOA calibration with different granularity due to different statistics available:
%Channel/Chip/Module-level calibration
%Overall good agreement between different layers/chips/channels
%Sometimes noise affects the reconstruction -> masking noisy channels

Due to the finite shaping time, the TOA measurement is affected by timewalk, i.e. an energy-dependence of the time measurement. For low-energy hits the timewalk reaches about 10\,ns.
In order to calibrate for the timewalk, the response-corrected TOA is compared to the MCP-PMT external time reference (Fig.~\ref{fig:toa_calib2}, right) and fitted with this function: $a + b x + c/(x-d)$.
The large fit residuals seen around 200\,MIP correspond to the switching of the ASIC gain, highlighting the importance of the gain inter-calibration.

%Time walk is considerable for low energy hits: about 15 ns in full E range > need to calibrate
%Using either the MCP as a reference, or a second channel with high signal (no TW)
%Fit function: a + b*x + c / (x -d)
%Gain transitions visible in fit residuals  —> inter-calibration is important!
%“Plateau” of distribution (a) gives systematic offset to reference
%Need to be taken into account when combining channels
%Time difference due to clock distribution within PCB clearly visible

\begin{figure}[b]
	\centering
	\includegraphics[width=0.4\linewidth]{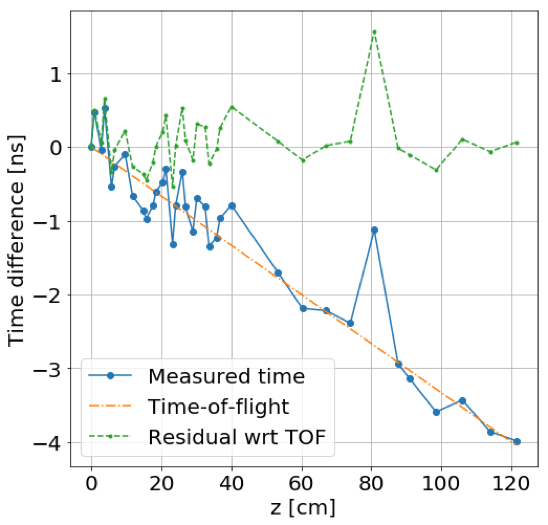}
	\includegraphics[width=0.58\linewidth]{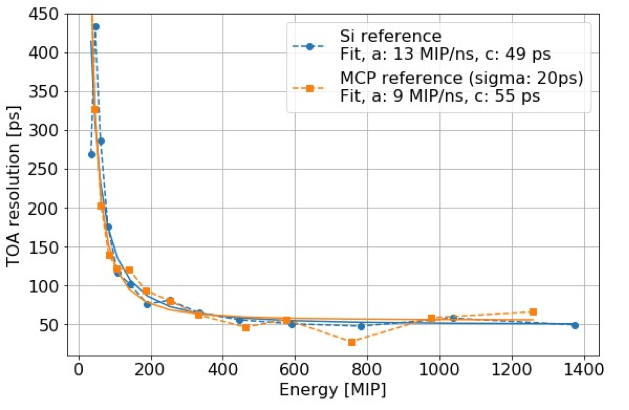}
	
	\caption{Left: difference of the measured TOA time with respect to the MCP-PMT reference in pion data. The trend follows the time-of-flight of pions and deviations are due to residual electronics effects.
		Right: single-channel time resolution using another silicon cell or the MCP as a reference  as described in the text. }
	\label{fig:toa_results}
\end{figure}

The constant term in the timewalk function $a$ provides the systematic offset to the time reference. Figure~\ref{fig:toa_calib2} (right) shows the offsets as determined for each central module of the setup: a clear trend corresponding to the time-of-flight for relativistic particles is visible  (Fig.~\ref{fig:toa_results}, left).
Since the prototype's clock distribution system ensures equal trace lengths to each detector module, deviations can be attributed to various residual electronics effects. It must be noted that the improper accounting of the systematic offsets will significantly deteriorate the final resolution when combining individual channels.

Finally, the single channel timing resolution was evaluated using both MCP-PMT and another silicon channel as reference (Fig.~\ref{fig:toa_results}, right). For the latter case the resolution was estimated based on the time difference between two channels for which the energy was required to be in the same energy bin; the two channel's measurements were treated uncorrelated and therefore the resolution was divided by a factor of $1/\sqrt{2}$.
The resolution was found to be compatible between the two references, with the constant term reaching the specification of the SKIROC2-CMS. The noise term (as in Sec.~\ref{sec:reco}) is degraded with respect to the single-chip lab measurements due to the increased noise environment in the prototype modules.

%Time resolution with MCP and independent channel as reference
%Reaching 50 ps constant term -> within SK2cms specification!
%Noise term worse than for single chip, due to increased (total and common) noise environment in the beam test
%Combining channels challenging (requires full calibration and precise offsets), but promising in improving shower timing ~√Nhit

\section{Outlook}

The present CMS endcap calorimeters will be replaced by the High Granularity Calorimeter for the HL-LHC era.
In order to improve  the physics potential, including pileup mitigation, HGCAL will provide timing information at the single-hit level.
%Simulation-based studies show a promising improvement from timing usage for the reconstruction. 
Clusters with $p_T >5$\,GeV will be able to achieve resolutions of at least 30\,ps.
%Photon/K0L resolutions down to 20/30 ps for pT > 2/5 GeV 

The electronics challenges of high-precision timing side are tackled by the front-end ASIC HGCROC using a multi-stage TDC. Preliminary measurements show a resolution below 13\,ps in a current prototype with a new iteration planned in 2020.

%Challenging front-end electronics: HGCROC with multi-stage TDC for TOA
%Resolution < 13 ps in current prototype, new ROC iteration to come

Beam tests have been performed with HGCAL prototypes in order to verify the technical and physics feasibility of timing measurements.
Data-driven TOA calibrations developed, and the single-channel resolution was found to be within the expectation of the SKIROC2-CMS prototype ASIC.
Further analysis, in particular the challenging combination of multiple channels, is ongoing. 
The results will be documented in a future publication.

%Precision timing at will be crucial  at the HL-LHC and very challenging for detector systems.

%\acknowledgments
%Support by Marie Curie?

\end{document}